\documentclass[twocolumn,showpacs,preprintnumbers,amsmath,amssymb]{revtex4}
\usepackage{color}
\usepackage{graphicx}
\usepackage{dcolumn}
\usepackage{bm}

\newcommand{\be}[1]{\begin{equation}\label{eq:#1}}
\newcommand{\ee}{\end{equation}}
\newcommand{\bea}{\begin{eqnarray}}
\newcommand{\eea}{\end{eqnarray}}

\begin{document}
\def\v#1{{\bf #1}}


\title{General Rule Predicting the Hidden Induced Order Parameters
and the Formation of
Quartets and Patterns of Condensates}

\author{Georgios Varelogiannis}
\affiliation{Department of Physics, National Technical University
of Athens, GR-15780 Athens, Greece}

\vskip 1cm
\begin{abstract}
We reveal the \emph{rule}
with which we predicted, and
verified by detailed numerical calculations on dozens of specific cases over the last decade,
the emergence of \emph{hidden}
induced states and the formation of
\emph{quartets} of order parameters.
The rule stipulates that if N order parameters are such that their
\emph{cyclic products} equal $\pm 1$  then the
presence of \emph{any} N-1 of them will necessarily induce the missing order
parameter as a \emph{hidden} order.
We demonstrate that the rule results from general microscopic
mean field theory and is thus universally valid.
Of exceptional physical interest are the cases N=4
called \emph{quartets}, and all $N>4$ cases decompose
into overlaping quartets which are
the building blocks of observable \emph{patterns of condensates}.
\emph{Quartets} may be found in two regimes, the \emph{hierarchy regime}
and the \emph{equity regime},
and transitions between the two regimes exhibit
universal characteristics.
Following the rule, we predict many characteristic examples of
\emph{quartets} involving all types
of particle-hole and superconducting condensates.
.
\end{abstract}

\maketitle

Mean field theory is a universal tool
in the study
of phase transitions.
The archetype of a microscopic mean field theory
for a quantum phase transition is
BCS theory of superconductivity \cite{BCS}
involving one order parameter.
With the discovery of unconventional superconductivity
emerged the need to consider more order parameters
enhancing the dimensions of the symmetry space \cite{SigristUeda,Volovik}.
Moreover, in many unconventional superconductors various types of
magnetic and/or charge ordered phases coexist or compete with superconductivity
and this requires a further enhancement
of the symmetry space \cite{SO(5)}.
However, with the exception of some early attempts
\cite{PsaltakisFenton,NassLevin}, the use of microscopic mean field theories for
multiple order parameters situations is avoided because of the rising complexity,
and more practical phenomenological Ginzburg Landau
schemes are adopted \cite{Littlewood}.

Here we reveal that in multiple order parameter spinor \emph{microscopic}
mean field theories,
there is an universal coupling that emerges between the parameters
that satisfy a simple rule, their \emph{cyclic} products equal either $+1$ or $-1$.
This interaction between order parameters is not anticipated from
group theory arguments and is unrelated with
details of the hamiltonians.
Order parameters that obey this (anti)unitary cyclic rule
aggregate into \emph{quartets} that represent
the building blocks of quantum complexity.
A \emph{quartet} is a set of \emph{four} order parameters
that may include the kinetic terms
of the Hamiltonian, that
are such that if any three of them are
present the fourth is necessarily present as well.
This is the generic mechanism through which
hidden induced phases emerge.

To demonstrate the relevance of the rule,
we have studied in detail dozens of specific
\emph{quartets} over the last decade.
We prove that, it is beyond any doubt,
that the above (anti)unitary cyclic product rule should be used \emph{apriori}
in any quantum spinor mean field theory
to make real physical predictions of great importance
that may affect the symmetry and even the topological characteristics
of the system under examination.
Moreover, qualitatively
new phenomena reported recently, like the sudden emergence of a \emph{quartet}
of coexisting condensates at high fields and low temperatures in
CeCoIn$_5$ \cite{PRLCeCoIn5}
reflect in fact global characteristics
of \emph{quartets} which may be tunned to switch from the
usual \emph{hierarchy} regime, where members of the
quartet are dominating, to the \emph{equity} regime
where all members manifest fully.
We provide characteristic \emph{quartets} that result from the
SU(8) Solomon and Birman theory
\cite{Birman} for particle-hole condensates and superconductivity.

We consider as an example
an eight dimensional spinor,
however our arguments are straightforwardly
generalized to higher dimensional spinors as well.
A mean field Hamiltonian is written as
$H =
\Psi^\dagger_{\bf k} \widehat{E}_{\bf k} \Psi_{\bf k}$
where $\Psi_{\bf k}$ are spinors
and
$\widehat{E}_{\bf k}$ is the $8\times 8$ energy
matrix
and the resulting Green's
function is:
\bea
\widehat{G}_o ( {\bf k} , i
\omega_n ) = {1 \over i \omega_n - \widehat{E}_{\bf k}}=\nonumber\\
=
%
 f_0 ( {\bf k},i \omega_n )
+ f_1 ( {\bf k},i \omega_n ) \widehat{E}_{\bf k} + ... + f_7 ( {\bf k},i \omega_n )
\widehat{E}_{\bf k}^7 \eea
Suppose now the Hamiltonian includes the order parameters $\bf{\alpha}$ and
$\bf{\beta}$ with corresponding matrices $\widehat{A}$, $\widehat{B}$,
and we examine if the order parameter $\gamma$ with matrix $\widehat{\Gamma}$
may be induced as a \emph{hidden} order.
All matrices are projected into a basis of tensor products of Pauli matrices,
$\widehat{\tau_i}\otimes\widehat{\rho_j}\otimes \widehat{\sigma_l}$
In the self-consistency equation that provides $\gamma$
\bea \gamma_{\bf k} =
T \sum_{\bf k'} \sum_n V^\alpha_{\bf k k'} {1 \over 8} Tr \{
\widehat{\Gamma} \widehat{G}_o ({\bf k'}, i \omega_n ) \} \eea
a non-zero trace exists only if a term proportional to the
matrix $\widehat{\Gamma}$ is generated and this can happen only if
$\widehat{A} \widehat{B} = c \widehat{\Gamma}$.
However, this contribution results from some power of $\widehat{E}_{\bf k}$,
and is \emph{always} accompanied by an additional contribution analogous to
$ \widehat{B}  \widehat{A}$. The condition for having
$\gamma$ induced as a hidden order from the
order parameters $\alpha$ and $\beta$  becomes $\{\widehat{A},\widehat{B}\} \neq 0$.
It is straightforward that we can only have
$\widehat{A}
\widehat{B} = \widehat{B} \widehat{A}$, or $\widehat{A} \widehat{B} =
- \widehat{B} \widehat{A}$ but the second case is irrelevant.
Therefore, $\gamma$ is induced as a hidden order if
$
\widehat{A} \widehat{B} = \widehat{B}\widehat{A}=c\widehat{\Gamma}$.
From that we get $\widehat{B} = c \widehat{A}\widehat{\Gamma}$ and
and $\widehat{B} = c \widehat{\Gamma}\widehat{A}$ and multiplying
the last two equations we obtain
$c = \pm 1$. While $c=1$ is anticipated, $c=-1$ is sometimes ignored even in this simple case.
With this, we now repeat the procedure in the other two order parameter channels,
and show that there is a term proportional to $\alpha \gamma$ in the
$\widehat{B}$ channel and one proportional to
$\beta \gamma$ in the $\widehat{A}$ channel of $\widehat{G}_o ( {\bf k} , i
\omega_n )$.
This implies that any one of $\alpha, \beta, \gamma$ will emerge necessarily
as a hidden order parameter if the
other two are present.
Such a \emph{trio} of coupled order parameters in which the coexistence of two implies the third
satisfies $ \widehat{A}\widehat{B}=\widehat{B}\widehat{A}=\pm\widehat{\Gamma}$, $\widehat{A}\widehat{\Gamma}=\widehat{\Gamma}\widehat{A}=\pm\widehat{B}$
and $\widehat{B}\widehat{\Gamma}=\widehat{\Gamma}\widehat{B}=\pm\widehat{A}$
which is equivalent to our cyclic rule:
\bea
\widehat{A}\widehat{B}\widehat{\Gamma}=\widehat{\Gamma}\widehat{A}\widehat{B}=
\widehat{B}\widehat{\Gamma}\widehat{A}=\pm 1
\eea

The physically relevant \emph{quartets}
emerge from a similar procedure, supposing that we have
three order parameters,
$\alpha$, $\beta$, $\gamma$, and we examine whether a hidden fourth order parameter
$\delta$ with corresponding matrix $\widehat{\Delta}$ can be induced:
$\widehat{A} \widehat{B}
\widehat{\Gamma} = c \widehat{\Delta} $.
As before, because this product may emerge by some power of $\widehat{E}_{\bf k}$
the necessary condition becomes
$\widehat{A} \widehat{B} \widehat{\Gamma} + \widehat{A} \widehat{\Gamma} \widehat{B}
+ ... \neq 0$
which is equivalent to
\bea
\{
\widehat{A} , \widehat{B} \} \widehat{\Gamma} + \{ \widehat{A} ,
\widehat{\Gamma} \} \widehat{B} + \{ \widehat{B}
, \widehat{\Gamma} \} \widehat{A} \neq 0\eea
In fact, one can prove that $\delta$ will be induced
if either one, or all three pairs of $\alpha$, $\beta$, and
$\gamma$ \emph{commute}. By repeating the procedure as before to the other
three order parameter channels, we conclude
that \emph{a quartet is made of four order parameters satisfying
the cyclic product rule}
\bea
\widehat{A} \widehat{B} \widehat{\Gamma}\widehat{\Delta}=
\widehat{\Delta} \widehat{A} \widehat{B}\widehat{\Gamma}=
\widehat{\Gamma} \widehat{\Delta} \widehat{A}\widehat{B}=
\widehat{B} \widehat{\Gamma} \widehat{\Delta}\widehat{A}=\pm  1
\eea

Similarly, we can generalize to
the case of four order
parameters inducing a hidden fifth:
$ \widehat{A} \widehat{B} \widehat{\Gamma} \widehat{\Delta} = c \widehat{E}$,
and the necessary and sufficient
conditions are now more complicated:
\bea \{\widehat{A},\widehat{B}\}
\{\widehat{\Gamma},\widehat{\Delta}\} + \{\widehat{A},\widehat{\Gamma}\}
\{\widehat{B},\widehat{\Delta}\} + \{\widehat{A},\widehat{\Delta}\}
\{\widehat{B},\widehat{\Gamma}\}+ \nonumber\\ \{\widehat{B},\widehat{\Gamma}\}
\{\widehat{A},\widehat{\Delta}\}  + \{\widehat{B},\widehat{\Delta}\}
\{\widehat{A},\widehat{\Gamma}\} + \{\widehat{\Gamma},\widehat{\Delta}\} \{\widehat{A},\widehat{B}
\} \neq 0 \eea
leading to a similar cyclic product rule.
However, this is useless
since one can show that the many order parameters states
decompose into a number of eventually overlapping quartets.
In fact, proceeding recursively, we can construct \emph{patterns of coexisting condensates}
involving more than one quartets as we will illustrate in a specific example.
This becomes more transparent,
and computationally more tractable for many order parameters, by noting
that the gap equation can be written as
\bea
\gamma_{\bf k}
=T
\sum_{{\bf k'},n} V^\gamma_{\bf k k'} {1 \over 8} { \lambda_2
\lambda_3 ... \lambda_8 + \lambda_1
\lambda_3 ... \lambda_8 + ... + \lambda_1 ... \lambda_7 \over
\lambda_1 \lambda_2 ... \lambda_8}
\eea
where $\lambda_i$ are the eigenvalues of $
\widehat{G}^{-1}_o ({\bf k'}, i \omega_n ) \widehat{\Gamma} $
which because $ Det [ \widehat{\Gamma} ] =
\pm 1 $ are in fact given by
\bea
Det \left[
\widehat{G}^{-1}_o ({\bf k'}, i \omega_n ) - \lambda \widehat{\Gamma} \right] = 0
\eea
Finally, we briefly note that the order parameters may be similarly mixed by
self-energy terms \cite{ImpuritiesOurs}.
For example, if we renormalize the propagator by
$\widehat{G} \approx \widehat{G}_o
+ \widehat{G}_o  \widehat{\Sigma} \widehat{G}_o
$
and take $ \widehat{\Sigma} \approx 
n U \widehat{\rho}_3 $, then
in the channel of $\alpha$
the product $\beta \gamma$ emerges if
$ \widehat{\Gamma} \widehat{\rho}_3 \widehat{B} =
\widehat{B} \widehat{\rho}_3 \widehat{\Gamma}$ or equivalently
$ \widehat{A} \widehat{B} \widehat{\Gamma} = \pm \widehat{\rho}_3 $.


\begin{table}[t]\caption{Inversion symmetric particle hole condensates. Multiply
by $\widehat{\rho}_3$ to obtain the odd in inversion counterparts (except Q) and note with tilde e.g.
$\widetilde{J^z_S} = \widehat{\tau}_2\widehat{\rho}_3\widehat{\sigma}_3$}
\begin{ruledtabular}
\begin{tabular}{|c|c|c|}\hline
Q &
$\widehat{\rho}_3 $ & particle hole asymmetry, chemical potential \\\hline
Y & $\widehat{\tau}_3 \widehat{\rho}_3 $ & nesting, particle hole symmetry, Pomeranchuk \\\hline
$S^x$ & $
\widehat{\rho}_3 \widehat{\sigma}_1 $ & conventional ferromagnet along x\\\hline
$S^y$ & $
\widehat{\sigma}_2 $ & conventional ferromagnet along y\\\hline
$S^z$ &  $ \widehat{\rho}_3
\widehat{\sigma}_3 $ & conventional ferromagnet along z\\\hline
$A^x$ & $\widehat{\tau}_3 \widehat{\rho}_3
\widehat{\sigma}_1 $ & d-wave ferromagnet, spin Pomeranchuk along x\\\hline
$A^y$ & $ \widehat{\tau}_3
\widehat{\sigma}_2 $ & d-wave ferromagnet, spin Pomeranchuk along y\\\hline
$A^z$ & $ \widehat{\tau}_3
\widehat{\rho}_3 \widehat{\sigma}_3 $ & d-wave ferromagnet, spin Pomeranchuk along z\\\hline
$C_Q$ & $
\widehat{\tau}_1 \widehat{\rho}_3 $ & conventional CDW\\\hline
$J_C$ & $ \widehat{\tau}_2 $ &
d-wave CDW, orbital antiferromagnet\\\hline
$S^x_Q$ & $ \widehat{\tau}_1
\widehat{\rho}_3 \widehat{\sigma}_1 $ & conventional SDW polarized along x\\\hline
$S^y_Q$ & $
\widehat{\tau}_1 \widehat{\sigma}_2 $ & conventional SDW polarized along y\\\hline
$S^z_Q$ & $
\widehat{\tau}_1 \widehat{\rho}_3 \widehat{\sigma}_3 $ & conventional SDW
polarized along z\\\hline
$J^x_S$ & $ \widehat{\tau}_2 \widehat{\sigma}_1 $ & d-wave SDW polarized along x\\\hline
$J^y_S$ & $ \widehat{\tau}_2 \widehat{\rho}_3
\widehat{\sigma}_2 $ & d-wave SDW polarized along y\\\hline
$J^z_S$ & $ \widehat{\tau}_2
\widehat{\sigma}_3 $ & d-wave SDW polarized along z\\\hline

\end{tabular}
\end{ruledtabular}
\end{table}
\indent

We have tested the real physical implications of our rule
on a
spinor space 
involving unconventional superconducting and particle-hole condensates,
many 
of which have been identified experimentally
in real material systems and are related to some of the most exciting phenomena
in many body physics and nanophysics. The 63 order parameters
involved are in fact generators of
the SU(8) spectrum generating algebra of Solomon and Birman \cite{Birman}.
Our discussion is based on the spinor \cite{PsaltakisFenton}
\begin{eqnarray}
\Psi^{\dagger}_{k}=\bigl(
c^{\dagger}_{k\uparrow},c^{\dagger}_{k\downarrow},
c_{-k\uparrow},c_{-k\downarrow},
c^{\dagger}_{k+Q\uparrow},c^{\dagger}_{k+Q\downarrow},
c_{-k-Q\uparrow},c_{-k-Q\downarrow}\bigr)\nonumber
\end{eqnarray}
that accounts for translation transformations over a commensurate
wavevector $\bf{Q}$, space inversion and spin rotations
on which the matrices
$\hat{\tau}_i,\hat{\rho}_j$ and $\hat{\sigma}_l$ act respectively.
For our argumentation the full palette of order parameters is necessary and is
reported on Tables I, II, and III. The eventual use of an alternative spinor
\cite{WeiMinZhang} would imply the
interchange between $\widehat{\tau}$ and $\widehat{\rho}$.

From the inversion symmetric order parameters of
Table I,
with the exception of Y, Q, $C_Q$ and $J^{x,y,z}_S$, all the others
break time reversal invariance.
By \emph{conventional} we mean \emph{any condensate that is even in translation}
and this includes $d_{xy}$ states, whereas d-wave is only
indicative referring to the well known
$d_{x^2-y^2}$ states from high-T$_c$ cuprates,
however it is as well valid for \emph{any other
representation that is odd in translation.}
The coexistence of a $d_{xy}$ $C_Q$ with $J_C$
imply the break of both parity in two dimensions and time reversal,
forming thus a \emph{chiral} state that accounts for the anomalous Nernst phenomena in the
pseudogap of the cuprates \cite{KotetesPRL}. If
on the other hand the $d_{xy}$ $C_Q$ coexists with
$J^{x,y,z}_S$ they form an \emph{helical} state \cite{Chakravarty}
which is also a topologic state with higher angular momentum
differing essentially on the fact that edge currents of different
spin orientations flow in opposite directions. Clearly, the aggregation of condensates
into \emph{quartets} may affect \emph{the topologic characteristics} of the system.
Multiplying with $\widehat{\rho}_3$ all states of Table I except Q provides the corresponding
odd in inversion particle-hole condensates that we note with a tilde. For example
the odd in inversion spin Pomeranchuk state along z
is written $\widetilde{A}^z=\widehat{\tau}_3\widehat{\sigma}_3$.
Breaking of the inversion obliges the order parameters that were breaking
time reversal to restore it, and the others  to break it.
Indeed, now the time reversal symmetry is only
broken by $\widetilde{Y}$, $\widetilde{C}_Q$ and
and $\widetilde{J}^{x,y,z}_S$.
A general discussion of unconventional density waves can be found in
\cite{Nayak} and Pomeranchuk states were discussed in
\cite{Yamase, WuZhangFradkin}.

In Table II are given the SC condensates
that are even under inversion.
Unless specified by an index $\Im$, the SC condensates are real.
Only the imaginary partners that result replacing
$\widehat{\rho}_2$ by $\widehat{\rho}_1$ break time reversal invariance. The 10
real SC condensates that break inversion
result by multiplying the inversion symmetric SC condensates
with $\widehat{\rho}_3\overrightarrow{\widehat{\sigma}}$ and are given in Table III.
The imaginary partners are now produced by
changing $\widehat{\rho}_1$ in $\widehat{\rho}_2$ and break time reversal
invariance.
The
non-centrosymmetric superconductors \cite{SigristNonCentro}
are the natural playground of odd SC condensates.
However, the coexistence
of $p_x$ with $p^{\Im}_y$ is a \emph{chiral} state
proposed for SrRuO$_3$ \cite{SigristSrRuO3}
and LiFeAs \cite{Buchner,SmallQLiFeAs}, and
there are also proposals for odd topologic superconductivity
in Cu$_x$Bi$_2$Se$_3$ as well \cite{Fu}, all these materials
being centrosymmetric.

\begin{table}[t]\caption{Inversion symmetric real SC condensates. For
the imaginary partners replace $\widehat{\rho}_2$ by $\widehat{\rho}_1$
and note with index $\Im$.}
\begin{ruledtabular}
\begin{tabular}{|c|c|c|}\hline
s-SC & $\widehat{\rho}_2
\widehat{\sigma}_2$ & conventional real\\\hline
d-SC & $\widehat{\tau}_3 \widehat{\rho}_2
\widehat{\sigma}_2$ & d-wave real\\\hline
$\eta$-SC & $\widehat{\tau}_1 \widehat{\rho}_2
\widehat{\sigma}_2$ & conventional staggered real \\\hline
$\pi_x$-SC & $\widehat{\tau}_2 \widehat{\rho}_2
\widehat{\sigma}_3$ & d-wave staggered triplet real along x \\\hline
$\pi_y$-SC & $\widehat{\tau}_2 \widehat{\rho}_2$ & d-wave
staggered triplet real along y\\\hline
$\pi_z$-SC & $\widehat{\tau}_2 \widehat{\rho}_2
\widehat{\sigma}_1$ & d-wave staggered triplet real along z\\\hline
\end{tabular}
\end{ruledtabular}
\end{table}

\begin{table}[t]\caption{Real, odd in inversion SC condensates
For the imaginary partners replace $\widehat{\rho}_1$ by $\widehat{\rho}_2$
and note with index $\Im$.}
\begin{ruledtabular}
\begin{tabular}{|c|c|c|}\hline
$\widetilde{s}_x$-SC & $\widehat{\rho}_1
\widehat{\sigma}_3$ & triplet even translation real along x\\\hline
$\widetilde{s}_y$-SC & $\widehat{\rho}_1$ & triplet even translation real along y\\\hline
$\widetilde{s}_z$-SC & $\widehat{\rho}_1
\widehat{\sigma}_1$ & triplet even translation real along z\\\hline
$p_x$-SC & $\widehat{\tau}_3 \widehat{\rho}_1
\widehat{\sigma}_3$ & p-wave real triplet along x\\\hline
$p_y$-SC & $\widehat{\tau}_3 \widehat{\rho}_1$ &
p-wave real triplet along y \\\hline
$p_z$-SC & $\widehat{\tau}_3 \widehat{\rho}_1
\widehat{\sigma}_1$ & p-wave real triplet along z \\\hline
$\widetilde{\eta}_x$-SC & $\widehat{\tau}_1 \widehat{\rho}_1
\widehat{\sigma}_3$ & staggered even translation real along x\\\hline
$\widetilde{\eta}_y$-SC & $\widehat{\tau}_1 \widehat{\rho}_1$ &
staggered even translation real along y\\\hline
$\widetilde{\eta}_z$-SC & $\widehat{\tau}_1 \widehat{\rho}_1\widehat{\sigma}_1$ &
staggered even translation real along z\\\hline
$\widetilde{\pi}$-SC & $\widehat{\tau}_2 \widehat{\rho}_1\widehat{\sigma}_2$ &
staggered singlet odd translation real\\\hline
\end{tabular}
\end{ruledtabular}
\end{table}

Having the full palette of order parameters, it is easy to exploit
the present rule 
finding a very large number of \emph{quartets}.
Note that members of \emph{quartets} may be 
the kinetic terms of the hamiltonian Y and Q, 
as well as the ferromagnetic order parameters $S^{x,y,z}$
that are equivalent to an applied Zeeman field. To produce
the quartets we are interested on we proceed as follows:
we start with a pair of condensates relevant to the considered problem,
then according to our rule,
they will interact with a third condensate if and only if,
the three pairs that these three condensates can form either all commute
or only one of the three pairs commutes. Then it is sufficient to
make the product of the there condensates to identify the fourth
with which they form a quartet. The correct mean field approach to the
problem should include \emph{all} four members of the quartet.
In Table IV are reported examples of quartets involving
only the even in inversion order parameters that are generators of an
SO(8) algebra and in Table V some quartets that involve a pair of
even and a pair of odd under inversion order parameters.

\begin{table}[t]\caption{Examples of quartets with inversion symmetric order parameters}
\begin{ruledtabular}
\begin{tabular}{||c||c||}\hline
Quartets & Quartets \\\hline\hline
Y,  Q,  $A^{x,y,z}$,  $S^{x,y,z}$ &
Y,  Q,  s-SC,  d-SC \\\hline
Q, $S^{x,z}_Q$, $A^{z,x}$, $J^y_S$ &
Q, $C_Q$, $S^{x,y,z}_Q$, $S^{x,y,z}$ \\\hline
Q, $J_C$, $J^{x,y,z}_S$, $S^{x,y,z}$ &
Q, $C_Q$, s-SC, $\eta$-SC \\\hline
Q, $\pi_y$-SC, $\pi_{x,z}$-SC, $S^{z,x}$ &
Q, s-SC, $\pi_y$-SC, $J^y_S$\\\hline
Y, d-SC, $\pi_y$-SC, $J^y_S$ &
Y, s-SC, $\pi_{x,z}$-SC, $S^{x,z}_Q$ \\\hline
Q, d-SC, $\pi_{x,z}$-SC, $S^{x,z}_Q$ &
Q, $\eta$-SC, $\pi_{x,z}$-SC, $A^{x,z}$ \\\hline
Y, $\eta$-SC, $\pi_{x,z}$-SC, $S^{x,z}$ &
Y, $C_Q$, $\eta$-SC, d-SC \\\hline
Y, $C_Q$, $A^{x,y,z}$, $S^{x,y,z}_Q$ &
Y, $S^{x,y,z}_Q$, $J^{y,z,x}_S$, $S^{z,x,y}$ \\\hline
Y, $S^{x,y,z}_Q$, $J^{z,x,y}_S$, $S^{y,z,x}$ &
$S^x$, $S^y$, $S^x_Q$, $S^y_Q$ \\\hline
$S^x$, d-SC, $\pi_{z}$-SC, $S^y_Q$ &
$S^{x,y,z}$, d-SC, $\pi_{x,y,z}^{\Im}$-SC, $C_Q$ \\\hline
$S^{x,y,z}$, d$^{\Im}$-SC, $\pi_{x,y,z}$-SC, $C_Q$ &
$S^{x,y,z}$, s-SC, $\eta$-SC, $S^{x,y,z}_Q$ \\\hline
$S^{x,y,z}$, s-SC, $\pi_{x,y,z}$-SC, $J_C$ &
$S^{x,y,z}$, s-SC, d-SC, $A^{x,z}$ \\\hline
$S^z_Q$, $C_Q$, $\pi_x$-SC, $\pi_y^{\Im}$-SC &
$S^z_Q$, $C_Q$, $\pi_x^{\Im}$-SC, $\pi_y$-SC \\\hline

\end{tabular}
\end{ruledtabular}
\end{table}

\begin{table}[t]\caption{Quartets mixing even and odd in inversion
OPs.}
\begin{ruledtabular}
\begin{tabular}{||c||c||}\hline
Quartets &  Quartets \\\hline\hline
Y, $S^{x,y,z}$, $\widetilde{Y}$, $\widetilde{S}^{x,y,z}$ &
Y, $S^{x,y,z}_Q$, $\widetilde{Y}$, $\widetilde{S}^{x,y,z}_Q$ \\\hline
Y, $A^{x,y,z}$, $\widetilde{Y}$, $\widetilde{A}^{x,y,z}$ &
Y, $C_Q$, $\widetilde{Y}$, $\widetilde{C}_Q$ \\\hline
Y, $J_C$, $\widetilde{Y}$, $\widetilde{J}_C$ &
Y. $J^{x,y,z}_S$, $\widetilde{Y}$, $\widetilde{J}^{x,y,z}_S$ \\\hline
Q, $S^{x,y,z}$, $\widetilde{Y}$, $\widetilde{A}^{x,y,z}$ &
Q, $A^{x,y,z}$, $\widetilde{Y}$, $\widetilde{S}^{x,y,z}$ \\\hline
$S^{x,y,z}$, $S^{z,x,y}_Q$, $\widetilde{Y}$, $\widetilde{J}^{y,z,x}_S$ &
Y, $S^z$, $p_x$-SC, $\widetilde{s}_y^{\Im}$-SC  \\\hline
Y, $S^z$, $p_x^{\Im}$-SC, $\widetilde{s}_y$-SC &
Y, $A^z$, $p_x$-SC, $p_y^{\Im}$-SC \\\hline
Y, $A^z$, $p_x^{\Im}$-SC, $p_y$-SC &
Y, d-SC, $\widetilde{A}^{x,y,z}$, $p_{x,y,z}$-SC \\\hline
Y, $C_Q$, $p_{x,y,z}$-SC, $\widetilde{\eta}_{x,y,z}$-SC &
Q, $S^z$, $p_x$-SC, $p_y^{\Im}$-SC \\\hline
Q, $A^z$, $\widetilde{s}_x$-SC, $p_y^{\Im}$-SC &
Q, s-SC, $\widetilde{A}^{x,y,z}$, $p_{x,y,z}$-SC \\\hline
Q, Y, $p_{x,y,z}$-SC, $\widetilde{s}_{x,y,z}$-SC &
Q, d-SC, $\widetilde{S}^{x,y,z}$, $p_{x,y,z}$-SC \\\hline
$S^{x,y,z}$, s-SC, $\widetilde{C}_Q $, $p_{x,y,z}$-SC &
$S^{x,y,z}_Q$, d-SC, $\widetilde{C}_Q$, $p_{x,y,z}$-SC \\\hline
$C_Q$, s-SC, $\widetilde{J}^{z,x,y}_S$, $p_{x,y,z}$-SC &
$C_Q$, d-SC, $\widetilde{A}^{x,y,z}$, $p_{x,y,z}$-SC \\\hline

\end{tabular}
\end{ruledtabular}
\end{table}

Remarkably, almost all interesting quartets
that we report in Table IV \emph{involve orders that break translational invariance}.
Unless is specified $\Im$ in a SC order of a quartet,
it is
understood that the quartet is valid for both the real and the imaginary
SC orders.
Note that a pair of order parameters belongs to more than one quartets.
Indeed the quartet (Q, $C_Q$, s-SC, $\eta$-SC) with the quartets (Q, $J_C$, d-SC, $\eta$-SC),
(Y, Q, s-SC, d-SC) and (Y, $C_Q$, d-SC, $\eta$-SC), have in common
two order parameters. We can form a closed \emph{pattern of condensates} out of these quartets
(Y, Q, $C_Q$, $J_C$, s-SC, d-SC, $\eta$-SC). Considering the real and imaginary parts of
the SC order parameters we obtain 10 order parameters that can be generators of a singlet
SO(5) model potentially relevant for high-T$_c$ cuprates. We will present in a following
publication with collaborators, how the particle-hole condensates in a pattern
of condensates could
help the emergence of pairs of SC condensates enhancing the critical
temperature.

The examples of \emph{quartets} reported in Tables IV and V involve
order parameters associated with virtually
all relevant phenomena in many body physics
and their potential implications will be discussed elsewhere.
We focus only to what we call the \emph{fundamental quartet}: chemical potential Q,
conventional
CDW and SDW, and FM ($Q, C_Q, S^{x,y,z}_Q, S^{x,y,z}$).
This quartet carries a \emph{profound and universal link
between charge and spin degrees of freedom}.  
Fully solving the spinor
mean field theory that corresponds in \cite{PRL2000} led us immediately to
the understanding of the present rule.
The first hints of this quartet were
in the \emph{excitonic ferromagnetism}
picture of Volkov and Kopaev \cite{VolkovKopaev} proposed
by Zhitomirsky, Rice and Anisimov
to explain weak ferromagnetism in
hexaborides \cite{Zhitomirsky}, however, Barzykin and Gorkov challenged this approach
identifying correctly that the charge would form a density wave \cite{Gorkov}.
Here we insert this phenomenon into the global context of \emph{quartets}
that enables a deeper and organized understanding of correlations between quantum ordered states.
Finally we simply note that on Table V one can easily identify quartets relevant for the
production of topologically non trivial states and quartets
potentially related with multiferroicity phenomena.

\begin{figure}[t]
\includegraphics[scale=0.60]{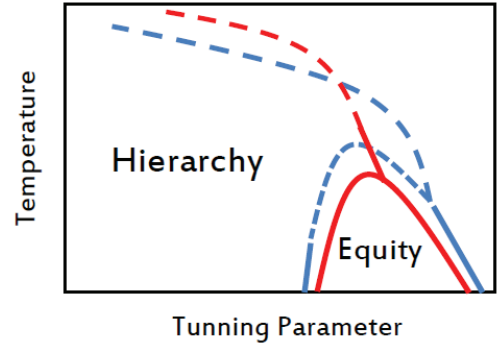}
\caption{Schematic description of the tuning by a physical parameter
from the hierarchy to equity regime. 
In the red case there is only one dominating order in the hierarchy regime.
In the blue case a tiny hidden order is induced in the hierarchy regime as well.
Full lines correspond to sharp first order transitions.}
\end{figure}
We have completed detailed studies for virtually all types of \emph{quartets}
exhibited in Table IV including
numerical explorations of the influence of
external fields or doping on their behavior. These results will be published
in following papers with collaborators, related previous results
can be found in
\cite{PRLCeCoIn5,TsonisJoP,JoPpitriplet,TsonisPhD,AperisPhD}.
We have identified only one case in which the quartet \emph{cannot}
form (s-SC, d-SC, Y, Q)
because the involved integrals cancel. 
In all quartets studied, we obtain two regimes. The \emph{Hierarchy}
regime is the most common situation in which usually there is one
dominating order but we may obtain in some cases
subdominant tiny hidden induced orders
dragged up to the critical temperatures of the dominating order
thus exhibiting a behavior inconsistent with thermodynamic expectations
in a phase transition. In the \emph{Equity} regime 
obtained by tuning parameters all orders attain comparable values.
The \emph{equity} regime defines
domes delimited by first order
transitions at low-T preventing the critical point
of the "dominating" orders.
It is possible to induce a transition from one regime to the
other exhibiting universal characteristics similar to those of Figure 1.
We have induced the transition from one regime to the
other with doping or with applying a magnetic field.
An example of the \emph{hierarchy} to \emph{equity} transition
for the quartet (Q, d-SC, $S^z_Q$, $\pi_z$-SC) tunned by the magnetic field
and corresponding to the red phase diagram in Figure 1
can be found in \cite{PRLCeCoIn5} where it has been
associated with transitions in the SC state of CeCoIn$_5$. 
An example of hidden order that survives   
in the hierarchy regime corresponding to the blue phase diagram in Figure 1
can be found in figure 2b of \cite{AlexCMR}.
Domes around expected quantum critical points in many different material systems
are the places where specific quartets fully develop.
In a following publication with collaborators we discuss in detail
the \emph{hierarchy} to \emph{equity} transition on specific examples of quartets
that we associate with experimentally observed domes.




In conclusion, we have verified that a simple rule
resulting from general microscopic mean field theory, predicts hidden
order parameters and the formation of quartets of order parameters
and should be used apriori to predict the correct
set of parameters for any spinor mean field theory.
The quartet interactions between order parameters that we report are
independent of the details of the hamiltonians involved and
exhibit universal characteristics.
The quartets are the fundamental
building blocks from which more complicated patterns of condensates may emerge.
We believe our results should be considered in any field of physics where there is
need for a spinor formalism.

I am indepted to Peter Littlewood, Gil Lonzarich and Ben Simons for
numerous enlightening
and stimulating comments and discussions,
collaboration on various parts of the project and continuous encouragement.
I am also grateful to my PhD and diploma students for their
precious help during the period of the project:
A. Aperis, M. Georgiou, G. Giannopoulos,
P. Kotetes, S. Kourtis, G. Livanas, G. Roumpos and S. Tsonis.
Work has been supported by the PEBE program of NTUA.


\end{document}